# ELECTROSTATIC POTENTIAL AS A DESCRIPTOR OF ANTI-BACTERIAL ACTIVITIES OF SOME ANACARDIC ACID DERIVATIVES: A STUDY USING DENSITY FUNCTIONAL THEORY

Manish K. Tiwari and P. C. Mishra*

Department of Physics, Banaras Hindu University, Varanasi - 221005, India

## Abstract

Structures and minimum molecular electrostatic potential (MEP) distributions in anacardic acid and some of its derivatives have been studied by full geometry optimization at the M06-2X/6-31G(d,p), WB97XD/6-31G(d,p) and B3LYP/6-31G(d,p) levels of density functional theory (DFT) in gas phase as well as in DMSO and aqueous solutions. Solvent effect was treated employing the integral equation formalism of the polarizable continuum model. Effects of modifications of the C1-side chain on the minimum MEP values in various regions were studied. Minimum MEP values near the oxygen atoms of the C2-OH group, oxygen or sulfur atoms of the C1-attached urea or thiourea groups and above or below the ring plane considered to be involved in interaction with the receptor were used to perform multiple linear regression. Experimentally observed anti-bacterial activities of these molecules against *S. aureus* are thus shown to be related to minimum MEP values in the above mentioned regions. Among the three DFT functionals used in the study, the M06-2X functional is found to yield most reliable results. Anti-bacterial activities have been predicted for certain molecules of the class which need to be verified experimentally.

# 1. Introduction

Chemical compounds obtained from plants and used as drugs have played great roles in human life from the ancient times [1,2]. Many of such species are known to possess anti-oxidant and anti-bacterial activities [1-6]. Anacardic acid (AA) (Fig. 1a) is a salicylic acid (SA) (Fig. 1b) derivative having a saturated unbranched long side chain at the C6-position of the phenolic ring [7]. In nature, AA usually occurs along with its three other forms i.e. 6[80(Z)-pentadecenyl]salicylic acid, 6[80(Z),110(Z)-pentadecadienyl]salicylic acid, and 6[80(Z),110(Z),140-pentadecatrienyl]salicylic acid [8,9], the C6-side chains of which have one, two and three C=C bonds (mono-, di- and tri-ene groups), respectively. These three other forms of AA are denoted by AA(15:n) (n=1-3), where n represents the number C=C bonds and 15 represents the number of carbon atoms present in the C6-side chain [8,9]. Following this notation, AA can also be denoted by AA(15:0) since there is no C=C bond in its C6-side chain. In this work, we would denote AA as AA(15:0) consistently. The nut (true fruit) of cashew (*Anacardium occidentale L.*), also known as Anacardiaceae apple, is the most important source of anacardic acid [8-11]. AA(15:0) is also extracted from cashew nut shell liquid (CNSL) which is a byproduct of cashew industry consisting of about 46-65% AA(15:0), 15-31% cardol, 10-22% cardanol and traces of methyl-cardol [12,13]. The mango fruit is another source of AA(15:0).

Anacardic acid has many useful pharmaceutical properties. As an anti-oxidant, it suppresses the carcinogenic activity associated with lipid peroxidation of fatty acids which leads to a decrease in membrane fluidity and disruption of its structure and function, as well as initiation and promotion of certain cancers. SA shows a weak and broad anti-bacterial activity against several microorganisms but it is not usable as an antibiotic [14]. However, AA(15:0) has a highly enhanced anti-bacterial activity mainly against Gram-positive bacteria [14] e.g. *S. mutans*, *B. ammoniagenes*, *S. aureus*, *B. subtilis* and *P. acnes* [15,16]. AA(15:0) also shows activity against some Gram-negative bacteria e.g. *H. pylori* which causes gastritis but it is not effective against certain other Gram-negative bacteria e.g. *E. coli*. AA(15:0) has the unique capability of changing the membrane potential and pH gradient across liposomal membranes [17]. A series of investigations of anti-bacterial and other activities of natural products like AA(15:0) were carried out by Kubo and coworkers [18-22]. The bactericidal activity of totarol was found to be highly enhanced against *S. aureus* when used in combination with AA(15:0) [18]. Variation of activity of AA(15:0) against different bacteria was studied by changing the alkyl side chain length [19].

It was found that AA(15:0) with side chain having 10 carbon atoms was most active against *S. aureus* while that with the side chain having 12 carbon atoms was most active against *P. acnes, S. mutants and B. ammoniagenes* [19,20]. The relationship of structure and antibacterial activity of AA(15:0) was studied against the methicillin resistant *S. aureus* ATCC 33591 (MRSA) bacterial strain [21,22]. The bactericidal activity of methicillin against MRSA strains was found to be significantly enhanced in combination with AA(15:0) [21,22]. Reddy and coworkers have studied activities of urea and thiourea derivatives of AA(15:0) against Gram-positive and Gram-negative bacteria and found that most of the compounds had comparable activities to that of the standard drug ampicillin [23,24]. Different other pharmaceutical and biological properties of AA(15:0) derivatives have been studied by other workers [25-27].

Several electronic indices e.g. atomic charges, highest occupied and lowest unoccupied molecular orbital (HOMO, LUMO) energies, dipole moments, total energies, heats of formation, ionization potentials and electron affinities etc. are generally used for analysis of structure-activity relationship [28]. Atomic charges are not rigorously defined observable quantities and cannot be used to obtain molecular electrostatic potential (MEP) distributions accurately, particularly MEP minima. It is also not logically clear what roles the other above mentioned indices would play in respect of molecular activities. In the present study, we considered hydrogen bonding between AA(15:0) or its derivatives and bacterial receptors to be important in controlling activities of the molecules. Minimum MEP values near given sites in a series of molecules obtained using continuous electron density are well known to be reliable measures of their relative hydrogen bond accepting strengths [29-38]. However, it is not yet established if MEP can serve as a single important measure of molecular activities as drugs. A multiple linear regression of MEP values versus anti-bacterial activities was performed for AA(15:0) and some of its derivatives with the objective to examine this aspect. We would also examine dependence of results on the quality of MEP distributions obtained employing three different commonly used density functionals. We hope to be able to identify the most reliable approach among the three approaches to predict anti-biotic activities of the given class of molecules most reliably.

## 2. Computational details
Geometries of all the molecules included in this study were fully optimized in gas phase as well as in aqueous and dimethylsulfoxide (DMSO) solutions employing the B3LYP [39], M06-2X

[40] and WB97XD [41] functionals of density functional theory (DFT) [42] along with the 6-31G(d,p) gaussian basis set. The integral equation formalism of the polarizable continuum model (IEF-PCM) was employed to treat solvation in aqueous and DMSO solutions [43]. Genuineness of the optimized minimum total energy geometries was ensured in all the cases by vibrational analyses at all the three levels of theory in gas phase as well as in aqueous and DMSO solutions. All the calculated vibrational frequencies were found to be real. Molecular electrostatic potential (MEP) values on molecular surfaces were calculated using continuously distributed electron density corresponding to the electron density value 0.001 electron/Bohr$^3$. At a given point **r** near a molecule, MEP V(**r**) (in a.u.) is given by the equation

$$V(\mathbf{r}) = \sum_A \frac{Z_A}{|\mathbf{R}_A - \mathbf{r}|} - \int \frac{\rho(\mathbf{r}')}{|\mathbf{r} - \mathbf{r}'|} d\mathbf{r}'$$

where $Z_A$ and $R_A$ are atomic number and position vector of an atom A, respectively while ρ(**r**′) is the continuous electron density at a point **r**′. MEP and molecular electric field (MEF) are known to be reliable descriptors of long range intermolecular interactions e.g. hydrogen bonding [29-38]. An interesting new property of MEP maps called σ-hole which is involved in halogen bonding has recently been shown by Clark et al. [33]. All the calculations performed employing density functional theory were carried out using the Windows version of the Gaussian09 (G09W) suite of programs [44]. MEP maps were drawn and molecular structures and vibrational modes visualized using the GaussView program (version 5.0) [45]. Molecular structures were drawn using ChemDraw [46].

**3. Result and discussion**

Both SA and AA(15:0) are bactericidal agents having the same i.e. COOH and OH groups attached to their C1 and C2 sites respectively (Figs. 1a,b). Since AA(15:0) is much more potent as a bactericidal agent than SA [19], its enhanced activity with respect to that of SA can be ascribed to the chain attached to its C6-site. It has been found that AA(15:0) with the C6-side chain having three C=C bonds (tri-ene group) is more potent than those having no or less numbers of ene groups [19]. Certain experimental studies have shown that several different substitutions in place of the C1-carboxylic group of AA(15:0) (Fig. 2) result in molecules with enhanced anti-bacterial potencies [23,24]. The minimum MEP values near the different sites of the molecules are given in Tables 1-3. The MEP features of the molecules of Tables 1 and 2 are presented in Figs. 3 and 4 respectively.

**3.1. Structures and MEP features of AA (15:n) (n=0-3)**

Structures of AA(15:0) and salicylic acid (SA) are shown in Figs. 1a,b respectively. Atomic numberings are also given in these figures. MEP values near the oxygen atoms of the C2-OH group (O2) and the C=O group (O1′) of the C1-carboxylic group in SA and AA(15:n) (n=0-3), and near the C=C double bonds of the C6-side chain obtained at different levels of theory in gas phase, aqueous and DMSO solutions are presented in Table 1. The MEP surfaces of SA and AA(15:n) (n=0-3) in aqueous solution obtained at the M06-2X/6-31G(d,p) level are given in Fig. 3. The negative MEP values near the oxygen atom of the C2-OH group (O2) (Fig. 1a-e) become larger negative in going from gas phase to aqueous or DMSO solutions in the different cases and the magnitudes of those in aqueous media lie in the range 45.1 to 49.9 kcal/mol at different levels of theory (Table 1). The magnitudes of MEP values near the oxygen atom of the C=O group (O1′) of the C1-carboxylic group are also increased in going from gas phase to aqueous and DMSO solutions at all the levels of theory, and these are usually appreciably smaller, lying in the range 34.4 to 37.2 kcal/mol in aqueous media, than those near the oxygen atom of the C2-OH group (O2) (Fig. 1). Due to large negative MEP values near the oxygen atoms of the C2-OH group (O2) and the C=O group (O1′) of the C1-carboxylic group in AA (15:n) (n=0-3), these sites would be expected to be involved in interaction with the receptor as hydrogen bond acceptors.

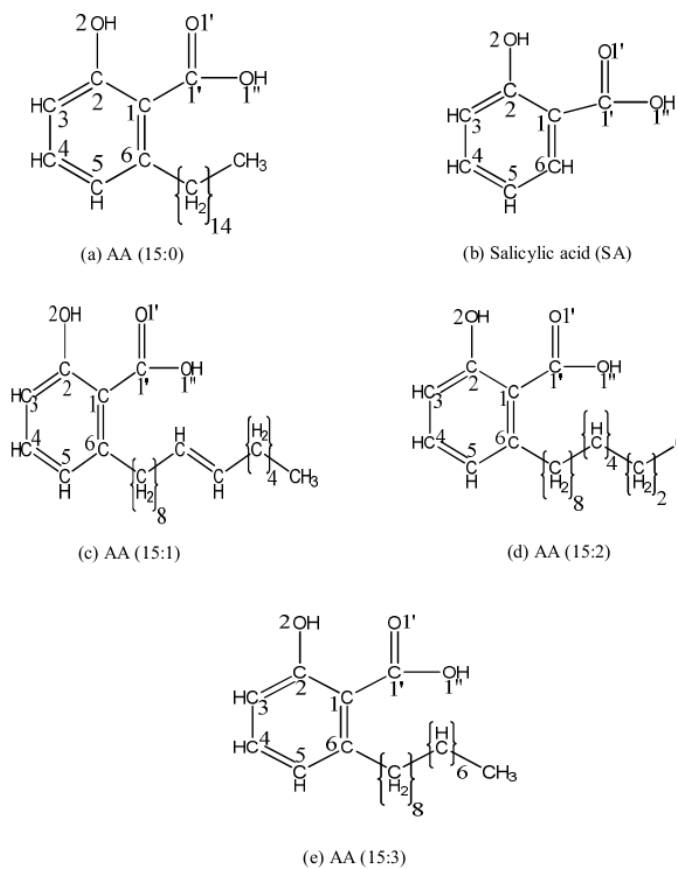

Fig.1. Structures of (a) Anacardic acid (AA) (15:0), (b) Salicylic acid, (c) AA (15:1), (d) AA (15:2) and (e) AA (15:3). MEP values near certain atomic sites of these molecules are presented in Table 1.

One, two and three ene-groups exist in AA(15:1), AA(15:2) and AA(15:3) molecules and MEP magnitudes near these groups lying in the range 22.2 to 29.2 kcal/mol obtained at different levels of theory in aqueous and DMSO media are found to become larger negative with increasing number of C=C double bonds (Table 1). However, these MEP magnitudes are appreciably smaller than those near the oxygen atoms O2 and O1′, and also these are increased in going from gas phase to aqueous or DMSO solution at all the levels of theory employed here (Table 1). Further, due to the C=C bonds, the side chain is progressively more bent in going from AA(15:0) to AA(15:3) (Fig. 3). Since the observed activities of these drugs in aqueous media follow the order [19] SA < AA(15:0) < AA(15:1) < AA(15:2) < AA(15:3) (Table 1), it appears quite likely that presence of the C6-side chain enhances potencies of drugs by facilitating their anchoring on the receptor surface.

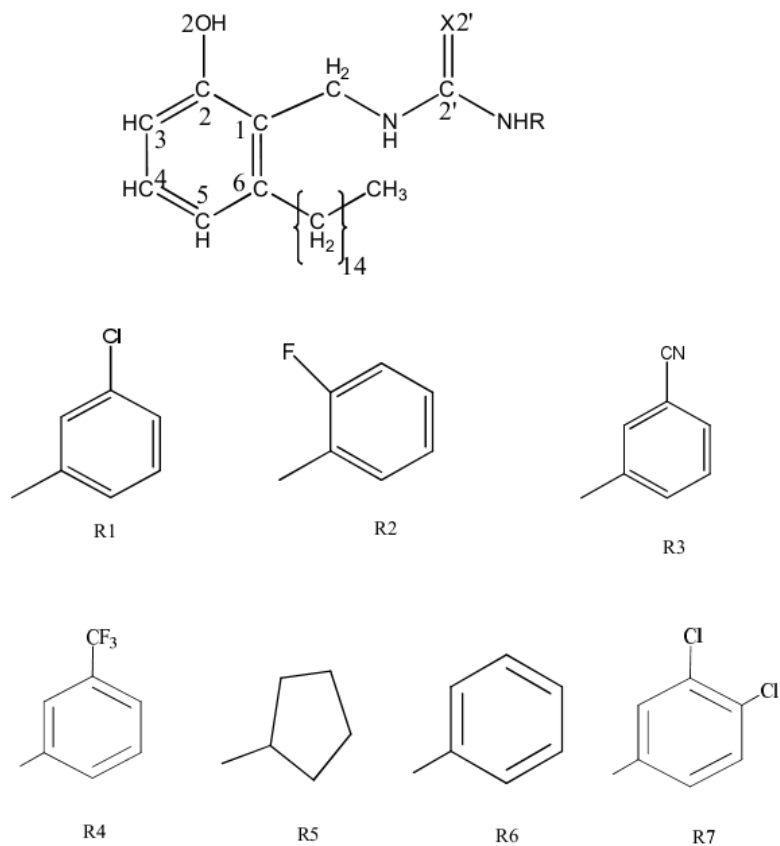

Fig. 2. Structures of urea (X2′=O) and thiourea (X2′=S) derivatives of AA (15:0). R has seven different structures R1-R7. When X2′=O and R=R1, R2 or R3, the molecules are labeled as A(Ur), B(Ur), and C(Ur), but when X2′=S and R=R1 or R3, the molecules are labeled as D(Th) and E(Th) respectively. MEP values near certain atomic sites of these molecules are presented in Table 2. The substituents R4 to R7 are present in molecules F(Ur), G(Ur), H(Ur) and I(Th) respectively in Table 3.

Experimentally observed minimum inhibitory concentrations (MIC) of SA and AA(15:n) (n=0-3) against *S. aureus* are given in Table 1. The linear correlation coefficients between the MEP values in the ene regions of three molecules i.e. AA(15:n) (n=1-3) obtained at the B3LYP/6-31G(d,p), M06-2X/6-31G(d,p) and WB97XD/6-31G(d,p) levels of theory and the corresponding $\log_{10}(1/\text{MIC})$ values are found to be 0.984, 0.974 and 0.964 respectively. These correlation coefficients indicate that the C6-side chain is involved in interaction with the receptor surface where MEP plays an important role.

**Table 1.** Minimum MEP values (kcal/mol) near the oxygen atoms of the C2-OH group (O2) and the C=O group (O1′) of the carboxylic group of salicylic acid (SA), AA (15:n) (n=0-3) and near the C=C double bonds (ene group) of the C6-side chain obtained at different levels of theory in gas phase, aqueous and DMSO solutions.[a]

| Molecule | Medium | B3LYP/6-31G(d,p) | | | WB97XD/6-31G(d,p) | | | M06-2X/6-31G(d,p) | | | Expt. Activity[c] MIC (μg/mL) |
|---|---|---|---|---|---|---|---|---|---|---|---|
| | | O2 | O1′ | Ene[b] | O2 | O1′ | Ene[b] | O2 | O1′ | Ene[b] | |
| SA | Gas | -38.5 | -33.3 | - | -39.4 | -33.9 | - | -39.4 | -32.9 | - | |
| | Aqueous | -45.1 | -36.6 | - | -46.2 | -37.2 | - | -46.1 | -36.1 | - | 400 |
| | DMSO | -45.1 | -36.6 | - | -46.1 | -37.1 | - | -46.0 | -36.1 | - | |
| AA (15:0) | Gas | -40.9 | -32.1 | - | -41.7 | -32.9 | - | -41.7 | -32.2 | - | |
| | Aqueous | -48.7 | -35.4 | - | -49.5 | -36.0 | - | -49.7 | -35.0 | - | >800 |
| | DMSO | -48.6 | -35.4 | - | -49.5 | -36.0 | - | -49.5 | -35.0 | - | |
| AA (15:1) | Gas | -40.8 | -32.4 | -20.5 | -41.7 | -32.3 | -22.3 | -40.2 | -32.7 | -20.9 | |
| | Aqueous | -48.6 | -35.5 | -22.2 | -49.5 | -35.2 | -24.3 | -49.9 | -34.4 | -23.8 | 100 |
| | DMSO | -48.5 | -35.5 | -22.2 | -49.4 | -35.2 | -24.2 | -49.7 | -34.4 | -23.8 | |
| AA (15:2) | Gas | -40.8 | -32.2 | -16.1 | -41.5 | -32.7 | -24.6 | -41.7 | -31.8 | -24.4 | |
| | Aqueous | -48.6 | -35.2 | -25.3 | -49.4 | -35.9 | -27.2 | -49.6 | -35.1 | -27.2 | 25 |
| | DMSO | -48.5 | -35.2 | -25.3 | -49.3 | -35.9 | -27.3 | -49.5 | -35.0 | -27.0 | |
| AA (15:3) | Gas | -40.9 | -32.4 | -24.0 | -41.7 | -32.8 | -25.9 | -41.7 | -32.1 | -26.0 | |
| | Aqueous | -48.8 | -35.5 | -26.8 | -49.4 | -35.8 | -29.2 | -49.7 | -35.0 | -29.2 | 6.25 |
| | DMSO | -48.6 | -35.4 | -26.7 | -49.3 | -35.7 | -29.1 | -49.5 | -35.0 | -29.2 | |

[a]Molecular structures are shown in Figs. 1(a-e) where atomic numbering is also given.
[b]Ene refers to the C=C double bond region in the C6-side chain. Mono-, di- and tri-ene forms of AA are AA(15:n), n=1-3, with one, two and three C=C bonds, respectively.
[c]Experimentally observed relative activities against S. *aureus* [Ref.19].

**3.2 MEP features of urea derivatives of AA(15:0)**

MEP values near the oxygen atoms of the C2-OH (O2) and C=O (O2′) groups as well as above or below the ring plane of AA(15:0) derivatives having different C1-attached urea derivative groups with varying substituents obtained at different levels of theory in gas phase, aqueous and DMSO solutions are presented in Table 2. The substituents include three (Cl, F, CN)-substituted phenyl groups beside X2′=O in urea derivatives and two (Cl, CN)-substituted phenyl groups beside X=S2′ in thiourea derivatives as shown in Fig. 2. It is found that in going from gas phase to aqueous or DMSO solution, in all the cases, MEP magnitudes near the oxygen atom of the C2-OH group (O2) are highly enhanced. In comparison to the MEP magnitudes in aqueous and DMSO solutions in AA(15:0), the enhancements of MEP magnitudes occurring due to the different substituents lie in the range ~26-42%. MEP surfaces of three urea derivatives of AA(15:0) (molecules A(Ur), B(Ur), C(Ur) of Table 2) in aqueous media obtained at the M06-2X/6-31G(d,p) level of theory are presented in Fig. 4. It is clearly seen that O2′ and O2 in all these molecules are associated with a large negative MEP region each. Instead of the C=O group of the C1-carboxylic group in AA(15:0), there is another C=O group in the C1-side chain in each of the molecules A(Ur), B(Ur) and C(Ur) (Fig. 2). If we examine the changes in MEP magnitudes near the oxygen atoms of these two types of C=O groups in going from AA(15:0) to the substituted derivatives A(Ur) and C(Ur) (Fig. 2, Table 2), we find that MEP magnitudes in each of these molecules are enhanced by about 17%, while in the case of B(Ur), the corresponding enhancement is ~82%. On the whole, we find that MEP magnitudes near the oxygen atom of the C=O group of the urea group (O2′) as well as above or below the ring plane are enhanced appreciably in various regions in going from AA(15:0) to its different urea derivatives.

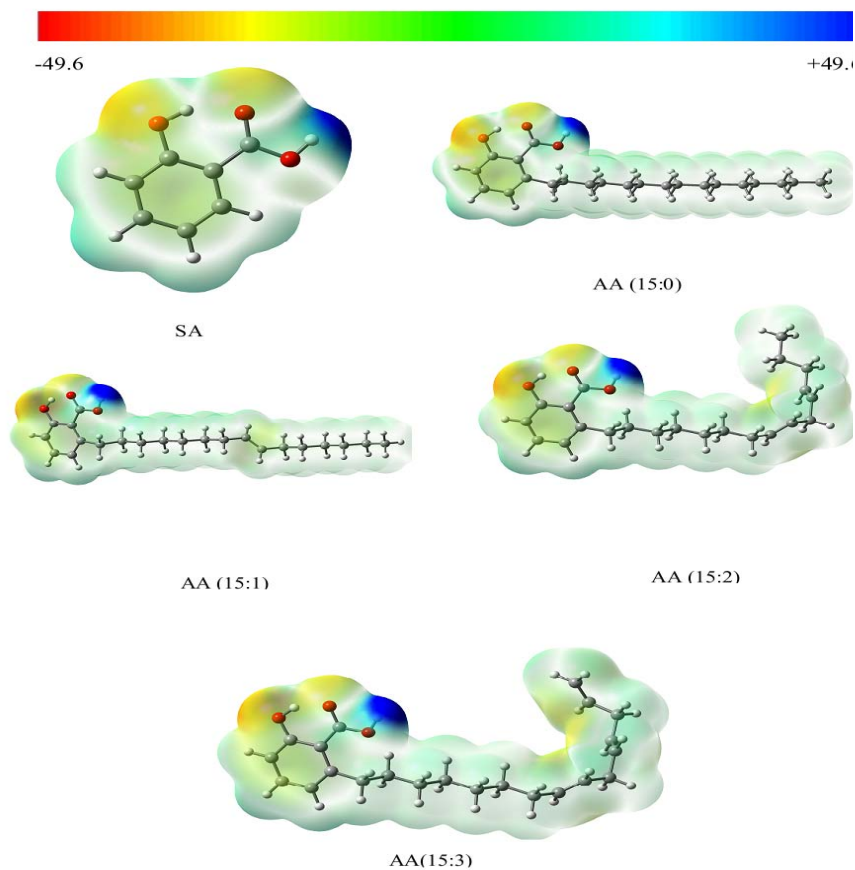

Fig 3. MEP maps (kcal/mol) of salicylic acid and different forms of anacardic acid (AA(15:n), n=0-3) on molecular surfaces corresponding to the electron density 0.001 electron/Bohr$^3$ obtained at the M06-2X/6-31G(d,p) level of theory in aqueous media. MEP values of these molecules are given in Table 1. MEP values follow the color code given at the top.

**Table 2.** Minimum MEP values (kcal/mol) near the oxygen atoms of the C2-OH group (O2) and C=O group (O2′), sulfur atom of the C=S group (S2′) and above or below the ring plane in AA (15:0) derivatives having C1-attached urea (A(Ur), B(Ur), C(Ur)) and thiourea (D(Th), E(Th))-substituted groups obtained at different levels of theory in gas phase, aqueous and DMSO solutions.[a]

| Mol-ecule | Medium | B3LYP/6-31G(d,p) | | | WB97XD/6-31G(d,p) | | | M06-2X/6-31G(d,p) | | | Expt. activity[b] |
|---|---|---|---|---|---|---|---|---|---|---|---|
| | | O2 | Ring | O2′/ S2′ | O2 | Ring | O2′/ S2′ | O2 | Ring | O2′/ S2′ | |
| A(Ur) | Gas | -50.7 | -20.6 | -35.5 | -52.3 | -22.1 | -38.1 | -52.2 | -21.7 | -35.3 | 88.9 |
| | Aqueous | -64.6 | -27.3 | -45.4 | -66.1 | -28.9 | -49.0 | -66.7 | -28.9 | -47.8 | |
| | DMSO | -64.5 | -27.1 | -45.2 | -65.6 | -28.9 | -48.3 | -66.4 | -28.7 | -47.8 | |
| B(Ur) | Gas | -54.7 | -23.3 | -49.4 | -55.4 | -23.7 | -50.2 | -55.6 | -24.5 | -50.0 | 94.4 |
| | Aqueous | -70.0 | -30.8 | -64.1 | -71.2 | -32.9 | -65.7 | -71.5 | -33.0 | -65.4 | |
| | DMSO | -69.8 | -30.6 | -63.9 | -70.9 | -32.7 | -65.6 | -71.2 | -32.8 | -65.1 | |
| C(Ur) | Gas | -49.6 | -19.2 | -31.9 | -51.0 | -20.8 | -31.5 | -51.0 | -20.5 | -33.2 | 83.3 |
| | Aqueous | -63.6 | -26.3 | -41.8 | -65.0 | -27.7 | -40.4 | -65.5 | -27.8 | -41.1 | |
| | DMSO | -63.4 | -26.0 | -41.6 | -64.6 | -27.5 | -40.1 | -65.3 | -27.7 | -41.0 | |
| D(Th) | Gas | -47.1 | -18.7 | -34.7 | -49.9 | -20.4 | -38.1 | -47.8 | -19.9 | -37.7 | 94.4 |
| | Aqueous | -60.8 | -24.2 | -51.7 | -65.1 | -26.8 | -55.1 | -64.8 | -31.3 | -52.8 | |
| | DMSO | -60.5 | -24.1 | -51.3 | -64.8 | -26.5 | -54.9 | -64.6 | -31.1 | -52.4 | |
| E(Th) | Gas | -47.5 | -17.1 | -34.2 | -49.0 | -19.8 | -37.7 | -48.8 | -19.8 | -34.8 | 100.0 |
| | Aqueous | -61.2 | -22.9 | -46.2 | -62.9 | -25.0 | -49.2 | -62.3 | -24.8 | -47.4 | |
| | DMSO | -60.9 | -22.8 | -46.0 | -62.5 | -24.9 | -49.0 | -62.1 | -24.7 | -47.3 | |

[a]Structures of the different urea (Ur) and thiourea (Th) derivatives are given in Fig. 2
[b]Experimentally observed percentage relative activities (PRA) with respect to ampicillin against S. *aureus* [Ref. 23,24].

### 3.3 MEP features of thiourea derivatives of AA(15:0)

MEP values near the oxygen atom of the C2-OH group (O2) and the sulfur atom of the C=S group (S2′) of the C1-attached thiourea group having different substituents in it (Fig. 2) obtained at different levels of theory in gas phase, aqueous and DMSO solutions are presented in Table 2. MEP surfaces of thiourea derivatives of AA(15:0) (molecules D(Th) and E(Th) of Table 2) in aqueous media obtained at the M06-2X/6-31G(d,p) level of theory are presented in Fig. 4. It is found that S2′ and O2 in all these molecules are associated with a large negative MEP region each. We observe that MEP magnitudes in these cases are different by small amounts from those presented in Table 2 where there is an oxygen atom (O2′) in place of the sulfur atom (S2′) as follows. First, the MEP magnitudes near the oxygen atom of the C2-OH group (O2) in the molecules D(Th) and E(Th) are different by not more than ~5% than those in the corresponding urea derivatives A(Ur), B(Ur) and C(Ur). Second, the MEP magnitudes near the sulfur

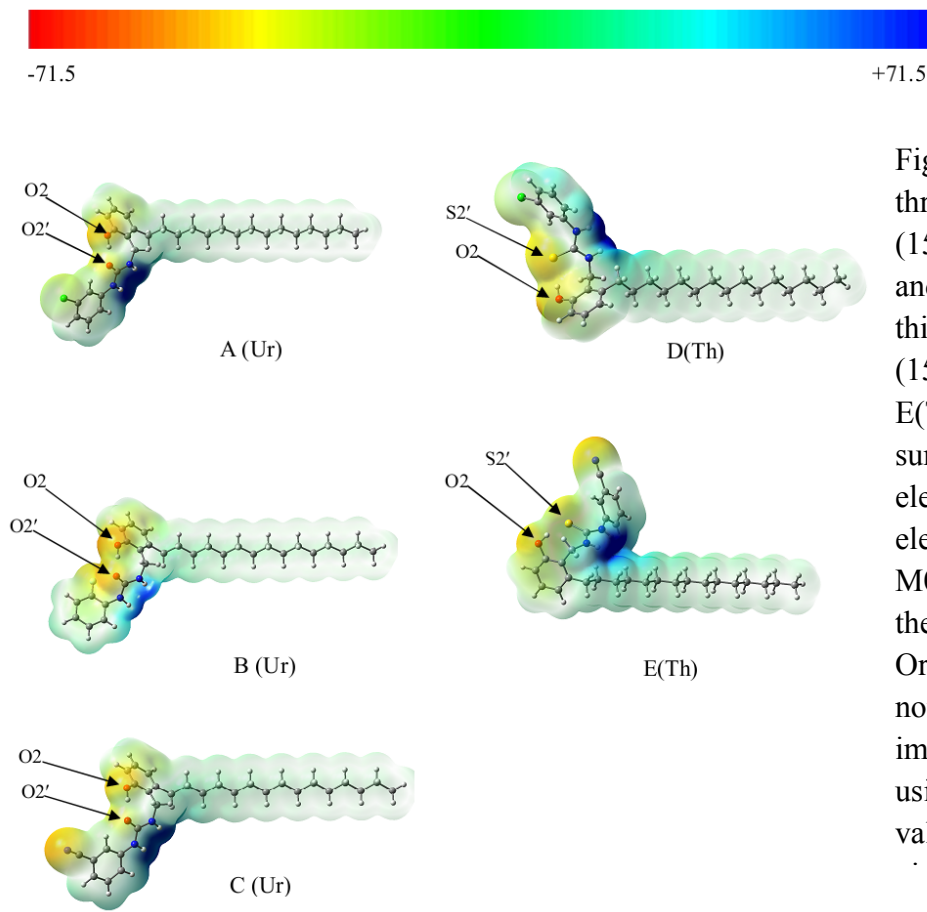

Fig 4. MEP maps (kcal/mol) of three urea derivatives of AA (15:0) (molecules A(Ur), B(Ur) and C(Ur) of Table 3) and two thiourea derivatives of AA (15:0) (molecules D(Th) and E(Th) of Table 3) on molecular surfaces corresponding to electron density value of 0.001 electron/Bohr$^3$ obtained at the M06-2X/6-31G(d,p) level of theory in aqueous media. Orientations of molecules are not the same in all cases and two important atoms are identified using arrows in each case. MEP values follow the color code

atom of the thiourea group (S2′) are enhanced by ~10% in comparison to those near the corresponding oxygen atom of the C=O group of each urea derivative. Third, in going from a urea group substitution to the corresponding thiourea group substitution, there is a decrease in MEP magnitude above or below the ring by ~10% in each case. The MEP magnitudes in the thiourea derivatives near the oxygen atom of the C2-OH group (O2) are larger by ~25% each in comparison to that in AA(15:0) (Table 1). Further, the MEP magnitudes near the sulfur atom of the C=S group (S2′) are larger by ~40% each in comparison to those near the oxygen atom of the C=O group (O1′) of the carboxylic group of AA(15:0).

## 3.4 Multiple linear regression using anti-bacterial activities and MEP values

In the experimental studies of anti-bacterial activities against *S. aureus*, certain urea derivatives have been found to be substantially more potent than AA(15:0) itself [23,24]. Let us define percentage relative activity (PRA) of such a derivative with respect to ampicillin against *S. aureus* as follows:

$$PRA = (P_M / P_A) \times 100 \qquad (1)$$

where $P_M$ = Activity of the given molecule and $P_A$ = Activity of ampicillin, both against *S. aureus*. Values of $P_M$ and $P_A$ are given in the literature [23,24] while those of PRA for the molecules included in this study are given in Table 2. Since the MEP values in aqueous and DMSO solutions generally differ only by less than 1% (Tables 2), multiple linear regression would be performed considering the MEP values in aqueous media only.

For multiple linear regression of urea and thiourea derivatives of anacardic acid, let us define the following:

Y= Molecular activity (PRA value),

X1= Magnitude of MEP near the oxygen atom of the C2-OH group (O2),

X2= Magnitude of MEP above or below the ring of a urea/thiourea derivative,

X3= Magnitude of MEP near the oxygen/sulfur atom (O2′/S2′) of the urea/thiourea group,

Now assuming linear dependence of Y on X1, X2 and X3, we have the equation

$$Y = a0 + a1X_1 + a2X_2 + a3X_3 \qquad (2)$$

where a0, a1, a2 and a3 are constants. We obtained values of these constants by multiple linear regression using the observed activities of five molecules (training set) and MEP magnitudes in aqueous media near the different sites as defined above at the different levels of theory (Table 3) as follows.

a0= 186.728, a1= -1.805, a2=-1.111, a3=1.125 (M06-2X/6-31G(d,p))

a0= 291.960, a1= -4.866, a2= 2.212, a3= 1.141 (WB97XD/6-31G(d,p))

a0= 10.746, a1= 2.958, a2= -5.252, a3=0. 605 (B3LYP/6-31G(d,p))

The magnitudes and signs of the regression constants obtained at the three levels of theory are quite different since the functionals M06-2X, WB97XD and B3LYP would deal with different interactions to different extents. The inclusion of dispersion interaction in the WB97XD hybrid functional plays an important role in studying interactions of aromatic rings with other molecules [41] while inclusion of an adequate percentage of Hartree-Fock exchange interaction in the M06-2X hybrid functional makes it very suitable for reliable prediction of kinetics, thermochemistry and non-covalent interactions [40]. The hybrid functional B3LYP predicts hydrogen bonding and several other molecular properties very well but it is not suitable to predict interactions involving aromatic rings [39]. Though we are not considering explicit interactions of the different molecules with the appropriate receptor, the above mentioned characteristics of the three functionals would be reflected in the relative MEP magnitudes near the various molecular sites, which would also affect the corresponding regression constants.

The PRA values for the five molecules A(Ur), B(Ur), C(Ur), D(Th) and E(Th) given in Table 2 in aqueous media can be predicted using the above mentioned regression constants at the different levels of theory. Standard deviations between the observed and predicted PRA values for these molecules (Table 2) at the M06-2X/6-31G(d,p), WB97XD/6-31G(d,p) and B3LYP/6-31G(d,p) levels of theory were found to be 0.61, 2.35 and 2.25 while the corresponding linear correlation coefficients were found to be 0.996, 0.940 and 0.946, the percentage levels of significance of these correlation coefficients being 0.025, 1.8 and 1.5 respectively. These data show that MEP values obtained at the M06-2X/6-31G(d,p) level of theory along with the regression constants are most suitable to predict PRA values for this class of molecules.

**3.5 Validation of regression: study of other urea and thiourea derivatives of AA(15:0)**

As noted earlier, the MEP magnitudes in aqueous and DMSO solutions differ generally by less than 1% (Tables 1,2). Therefore, we would present and discuss the MEP magnitudes in aqueous media only. MEP magnitudes near the oxygen atoms of the C2-OH group (O2) and C=O group (O2′), sulfur atom of the C=S group (S2′) and above or below the ring plane in AA (15:0) derivatives having C1-attached urea (F(Ur), G(Ur), H(Ur)) and thiourea (I(Th))-substituted groups obtained at different levels of theory in aqueous media are presented in Table 3. This table also contains a comparison of the predicted activities of the four molecules using the regression constants obtained as discussed above in Section 3.4 with the experimentally observed values [23,24]. The magnitudes of MEP presented in Table 3, like those of Table 2, follow the order O2 > O2′/ S2′ > (above or below the ring plane). The standard deviations between the theoretically predicted and experimentally observed activities at the M06-2X/6-31G(d,p), WB97XD/6-31G(d,p) and B3LYP/6-31G(d,p) levels of theory are found to be 2.7, 5.3 and 6.7 respectively. Thus we find that the theoretical predictions made using the M06-2X/6-31G(d,p) level of theory are consistently in best agreement with the experimentally observed ones.

**Table 3.** Minimum MEP values (kcal/mol) near the oxygen atoms of the C2-OH group (O2) and C=O group (O2′), sulfur atom of the C=S group (S2′) and above or below the ring plane in AA (15:0) derivatives having C1-attached urea (F(Ur), G(Ur), H(Ur)) and thiourea (I(Th))-substituted groups obtained at different levels of theory in aqueous media.[a]

| Molecule | Functional | MEP values | | | Activity(PRA)[b] | |
|---|---|---|---|---|---|---|
| | | O2 | Ring | O2′/ S2′ | Expt. | Predicted |
| F(Ur) | M06-2X | -66.4 | -28.1 | -50.0 | 94.4 | 91.9 |
| | WB97XD | -66.0 | -29.0 | -48.7 | | 90.5 |
| | B3LYP | -64.4 | -27.1 | -45.1 | | 86.2 |
| G(Ur) | M06-2X | -70.6 | -32.7 | -55.9 | 88.9 | 85.9 |
| | WB97XD | -69.4 | -32.5 | -58.0 | | 92.3 |
| | B3LYP | -68.3 | -31.1 | -58.4 | | 84.8 |
| H(Ur) | M06-2X | -67.8 | -30.1 | -54.1 | 88.9 | 91.8 |
| | WB97XD | -66.3 | -29.9 | -53.1 | | 96.1 |
| | B3LYP | -65.4 | -28.4 | -49.2 | | 84.8 |
| I(Th) | M06-2X | -62.0 | -24.9 | -46.8 | 100.0 | 99.8 |
| | WB97XD | -63.4 | -25.3 | -49.6 | | 96.0 |
| | B3LYP | -59.5 | -23.1 | -45.4 | | 92.9 |

[a]Structures of the urea (Ur) and thiourea (Th) derivatives are given in Fig. 2.
[b]Theoretically predicted and experimentally observed percentage relative activities (PRA) with respect to ampicillin against *S. aureus* [Ref. 23,24].

### 3.6 Predicted activities of new thiourea derivatives of AA(15:0)

Anti-biotic activities of some hydroxy, amino, methyl, fluoro and chloro-substituted derivatives of the molecule E(Th) of Table 2 were predicted using the M06-2X functional along with the 6-31G(d,p) basis set and the regression constants mentioned above. MEP values of these molecules are not presented and only the M06-2X functional was used here since it has been shown be most reliable in the present context. The predicted activities (PRA values) of these molecules given in Table 4 reveal that we should expect the ortho-OH derivative of E(Th) (Table 2) to be ~22% more active as a new anti-bacterial drug against *S. aureus* than ampicillin while the ortho-$NH_2$, meta-Cl and para-Cl derivatives would be expected to be more active than the same by 5-10%. Other derivatives would have activities comparable to that of ampicillin or less than the same. An experimental test of these predictions is needed.

| Substituent | Ortho | Meta | Para |
|---|---|---|---|
| OH | 122.1 | 98.6 | 88.3 |
| NH2 | 105.5 | 84.1 | 80.4 |
| CH3 | 93.6 | 84.7 | 86.3 |
| F | 85.4 | 101.4 | 100.6 |
| Cl | 84.8 | 109.7 | 104.6 |

**Table 4.** Predicted activities of different OH, $NH_2$, $CH_3$, F and Cl-substituted derivatives of the E(Th) molecule (Table 2) obtained using the M06-2X functional along with the 6-31G(d,p) basis set in aqueous media.[a]

[a]Structures of these molecules are shown in Fig. 2 where X2′= S and R= R3. Ortho, meta and para positions with respect to C2 are the positions C3, C4 and C5 in Fig. 2, respectively.

**Conclusion**

The main conclusion of this study is as follows. Multiple linear regression was performed considering linear dependence of the experimentally observed anti-bacterial activities of certain AA(15:0) derivatives against *S. aureus on* the MEP values near the oxygen atoms of the C2-OH group (O2), the atoms O2′/S2′ of C1-substituted urea/thiourea groups and those above or below the ring plane. Anti-bacterial activities of several molecules have been explained successfully using MEP as a logically meaningful and comprehensive electronic descriptor of molecular activities. The predicted values of known molecular activities of some molecules at the M06-2X/6-31G(d,p) level are in best agreement with experimental values. According to the predicted anti-biotic activities of certain new ortho, meta and para OH, $NH_2$, $CH_3$, F and Cl-substituted derivatives of a particular molecule (molecule E(Th) of Table 2), the ortho-OH derivative would be ~22% more active as a new anti-bacterial drug against *S. aureus* than ampicillin while the ortho-$NH_2$, meta-Cl and para-Cl derivatives would be more active than the same by 5-10%. Activities of the other derivatives would be comparable to that of ampicillin or less than the same. These predictions need to be tested experimentally.

**Acknowledgement**

One of the authors (PCM) is thankful to the National Academy of Sciences (NASI) for awarding a Senior Scientist Fellowship and for financial support. MKT is thankful to the University Grants Commission (New Delhi) for a research fellowship. Thanks are due to Dr. A.K. Saxena, Central Drug Research Institute, Lucknow for valuable discussion.